\documentclass{article}
\pdfoutput=1
\usepackage{microtype}
\usepackage{graphicx}
\usepackage{subfigure}
\usepackage{booktabs} 

\PassOptionsToPackage{hyphens}{url}\usepackage[hyperindex,breaklinks]{hyperref}

\usepackage{xcolor}

\usepackage[accepted]{icml2020}

\icmltitlerunning{The Limits of Preference Elicitation for Participatory Algorithm Design}

\begin{document}

\twocolumn[
\icmltitle{What If I Don't Like Any Of The Choices?\\
    The Limits of Preference Elicitation for Participatory Algorithm Design}



\icmlsetsymbol{equal}{*}

\begin{icmlauthorlist}
\icmlauthor{Samantha Robertson}{eecs}
\icmlauthor{Niloufar Salehi}{ischool}
\end{icmlauthorlist}

\icmlaffiliation{eecs}{Department of Electrical Engineering and Computer Sciences, University of California, Berkeley, California, USA}
\icmlaffiliation{ischool}{School of Information, University of California, Berkeley, California, USA}

\icmlcorrespondingauthor{Samantha Robertson}{samantha\_robertson@berkeley.edu}

\icmlkeywords{Matching algorithms, participatory algorithm design}

\vskip 0.3in
]



\printAffiliationsAndNotice{}  

\begin{abstract}
Emerging methods for participatory algorithm design have proposed collecting and aggregating individual stakeholder preferences to create algorithmic systems that account for those stakeholders' values. Using algorithmic student assignment as a case study, we argue that optimizing for individual preference satisfaction in the distribution of limited resources may actually inhibit progress towards social and distributive justice. Individual preferences can be a useful signal but should be expanded to support more expressive and inclusive forms of democratic participation.
\end{abstract}

Machine learning is  increasingly used in high-stakes situations such as access to education, employment, and healthcare \cite{Ghassemi2019, Raghavan2019}. Documented instances of discrimination, e.g. \cite{Buolamwini2018, Obermeyer2019, Chouldechova2017}, have led to mounting pressures to improve these systems by accounting for stakeholder values and context in their design and use. One proposed way to do so is to incorporate individual preferences into algorithmic decision-making. Some emerging\footnote{This relatively new approach in machine learning builds on a substantial body of work in social choice theory and economics that seeks to understand how individual preferences can be aggregated to inform decisions on behalf of a group. \cite{socialchoice} provides an overview.} methods for participatory algorithm design have proposed collecting and aggregating individual stakeholder preferences to create algorithmic systems that represent the values and goals of those stakeholders \cite{webuildai, Kahng19, Wang2020, Noothigattu2018}. 

While these approaches offer some channels for stakeholders to provide input, in this paper we argue that individual preference aggregation is insufficient to support participatory, democratic governance of algorithmic systems. We examine the student assignment algorithm\footnote{While this is not a machine learning algorithm, the widespread adoption of matching algorithms for student assignment over a number of years makes them a useful case study for understanding the implications of preference-based algorithmic systems more broadly.} in San Francisco Unified School District (SFUSD) as a case study. Incoming students in San Francisco submit a ranked list of schools they would like to attend, and the assignment algorithm matches students to schools in a way that optimally satisfies the students' preferences \cite{policy5101}. In practice, the algorithm has not lived up to its theoretical promises of transparency, efficiency, and fairness \cite{Abdulkadiroglu2003}. In fact, it is difficult for families to navigate and has exacerbated segregation, with students from historically marginalized backgrounds concentrated in under-served schools \cite{sfusd2018}. For these reasons the school district voted to stop using and completely redesign the system in 2018. Based on this case study, we problematize three assumptions that are implicit in preference-based algorithmic systems more broadly: (1) that each individual has some inherent, fixed preferences over the available set of alternatives; (2) that these preferences fully encapsulate each individual's relevant values, needs, and goals; and (3) that some aggregation of these preferences reflects what is best for the group collectively. 

This case study demonstrates that reporting preferences over a fixed set of alternatives is an insufficient and unequal way to give stakeholders a voice in algorithmic decision-making. Instead, preferences often reflect existing social biases and inequities and, for this reason, we should not assume that an aggregation of those preferences will promote equity and justice. Individual preferences can be a useful signal, but must be expanded to include more expressive and context-aware forms of democratic participation. For example, student assignment systems could support alternative formats for expressing priorities, provide more avenues to express preferences, or facilitate public discourse and deliberation regarding desirable social outcomes and relevant design decisions given technical and resource constraints.

\section{Algorithmic Student Assignment}

In most public school districts, students are assigned to schools based on where they live. As a result, racial segregation and economic inequalities are reflected in segregated and unequal schools. Increasingly, school districts have been introducing school choice systems that allow students to apply to schools across the district. Students submit a ranked list of schools they would like to attend and the district uses an algorithm to match students to schools based on those preferences. These policies have been met with great excitement for their ability to advance equitable access to high quality education, create more diverse classrooms, and provide flexibility to families \cite{Kasman2019}. New York City was the first to introduce a matching-based assignment system for high schools in 2003, and many school districts across the country have followed suit in the decades since \cite{Roth2015}. 

A number of matching algorithms have been developed by economists in the field of market design to find an optimal matching in a two-sided market based on each side's preferences \cite{GaleShapley1962, ShapleyScarf1974}. In student assignment, the two sides of the market are the incoming students and available seats in the schools. Students report their preferences by ranking the schools, and schools can define priority categories for students, such as priority for younger siblings of continuing students or priority for students living in the school's surrounding neighborhood. The matching algorithms used in the student assignment context\footnote{Deferred Acceptance \cite{GaleShapley1962} and Top-Trading Cycles \cite{ShapleyScarf1974} are commonly used for student assignment \cite{Abdulkadiroglu2003}.} are \textit{student-optimal} in the sense that they are optimized to satisfy student preferences as efficiently as possible,\footnote{For more details about properties of matching mechanisms and matchings, such as strategy-proofness, and trade-offs between stability and efficiency, see \cite{Abdulkadiroglu2003}. For the purposes of this paper, it is most important to keep in mind that the primary goal of these algorithms is to satisfy student preferences.} subject to each school's capacity constraints \cite{Abdulkadiroglu2003}. School priorities only determine the order in which students are offered over-demanded seats, with ties between students with the same priority resolved by a random lottery. 

\subsection{San Francisco Unified School District}

San Francisco Unified School District (SFUSD) introduced a student assignment system based on a matching algorithm\footnote{SFUSD appears to use Top-Trading Cycles.} in 2011 in the hopes of promoting equitable access to educational opportunity and diverse classrooms \cite{policy5101}. However, by 2018 the Board had voted to redesign the system in response to widespread dissatisfaction amongst families and clear evidence that the system was not serving the district's goals \cite{sfusd2018}. The district was especially concerned that the algorithm had been unable to promote diverse classrooms and equitable access to education, largely due to racial and socioeconomic segregation in families' preferences \cite{sfusd2019}. As part of an ongoing project, we have conducted an in-depth case study to understand why the promises of market design have not been realized in SFUSD. For this work we have conducted interviews with parents in San Francisco and performed content analysis on policy documents by the school district that describe the goals of the assignment algorithm. We use this system as a running example to demonstrate some of the problems that arise when using individual preferences to determine the distribution of public resources.

\section{What do I prefer?}

Theoretical market design literature typically assumes that each individual possesses some inherent preferences that represent the relative value they would gain from each of the available options. In practice, forming a preference list is time-consuming for families and is strongly shaped by social context. 

Families invest heavily in information acquisition, often seeking out more information than theoretical incentives would predict \cite{Chen2017}. In interviews with San Francisco families, we found that researching schools is especially time-consuming and frustrating. Families have a choice between all 72 elementary schools in the district and they are able to rank as many as they would like in their preference list. Many parents attend multiple in-person school tours and struggle to navigate fragmented and disorganized online resources. As a result, families require significant time and resources to fully understand the available options.

Since preferences are formed through a situated, \textit{ad hoc} process of information gathering, they reflect and replicate implicit biases and social context, rather than an objective measure of each school's value to the family. Experimental studies of advice sharing in the context of student assignment have revealed that information sharing in social networks strongly influences whether individuals truthfully report their preferences or engage in strategic behavior \cite{Ding2017, Ding2019}. In San Francisco, we found that parents frequently consult other parents in their network for information about schools, both online and in person. White, economically advantaged parents acknowledged a strong preference amongst their peers for a small subset of the available schools that are perceived as the ``best.'' Consistent with previous findings (e.g., \cite{Johnson2003}), these schools are often also those with more white, economically advantaged students \cite{collins}.

Systems that over-emphasize reported preferences as a form of participation risk essentializing these preferences as an objective, true representation of an individual's values and goals. Despite struggling to form their preference lists in the first place, many parents we spoke to staunchly defended their ultimate list as \textit{the} best reflection of their child's interests. Forms of participation that encourage stakeholders to essentialize and rationalize their own views risk legitimizing damaging biases and stereotypes, and making trade-offs between stakeholders' interests increasingly difficult. For example, SFUSD's efforts to broaden access to education have faced push-back from advantaged families who are intent on sending their child to a heavily over-demanded school \cite{collins}.

\section{What if I don't like any of the choices?}

Matching algorithms for student assignment also implicitly assume that preferences provide \textit{all} relevant information about those families' values and needs pertaining to school choice. On the contrary, we argue that preferences can only represent a subset of families' possible values and needs, and furthermore that preferences are a more expressive form of participation for those who have historically benefited more from and had more control over the education system.

In many contentious policy domains like student assignment, the set of alternatives that individuals can rank has been shaped by structures of power and inequality.  By allowing students to apply to any school in the district, SFUSD appears to offer equal access to educational opportunities, but the same set of alternatives does not offer identical opportunities to all students. Most families want a school that is close to home and that will provide their child with a high-quality education \cite{Burgess2015}. However, prior studies have shown that families with more resources are able to more heavily prioritize academic factors, while other families may be constrained in their choice of schools by economic and social factors \cite{Laverde2020, Hastings2009}. Similar patterns of access exist in San Francisco. Through interviews, we found that a lack of school bus routes constrains the available schools for parents who do not have access to a car or time to commute with their child. Further, the African American Parent Advisory Council (AAPAC) has pointed out that African American students in San Francisco continue to face racism and biases in racially diverse classrooms \cite{aapac}.  The literature on matching algorithms focuses on efficiently satisfying preferences without acknowledging the ways in which this emphasis disempowers those who cannot or do not want to attend the schools that are currently more well-funded and provide more opportunities \cite{Scott2013}.

There are various policy approaches to educational equity that are not based on school choice or student assignment, but preferences provide a very limited avenue for families to advocate for these types of reforms.  For example, the AAPAC has criticized the SFUSD system for pushing under-served students out of their communities in order to access high quality schools. They have called on SFUSD to create high quality programs in low-resource neighborhoods and protect local students’ access to these programs. Since submitting preferences is the only way many parents interact with the enrollment system and communicate their needs to the district, support for these demands may not be fully received. Even when other avenues for participation are available, forming preferences is already extremely time-consuming and many families may not have the time and resources to engage further with the district.

\section{What do \textit{we} prefer?}

The limitations of preferences explored in the previous two sections are exacerbated when preferences are aggregated to make decisions that impact a group. Assigning students to public schools determines which children are given access to a well-funded, high-quality education, making questions of distributive justice extremely relevant. Principles of distributive justice concern how a group should allocate limited resources among members with competing needs \cite{Roemer1996}. Preference aggregation procedures assume that the ideal distribution is the one in which every individual receives their first choice. For student assignment, capacity constraints and uneven demand usually make this ideal outcome impractical, but trade-offs are made to satisfy preferences as efficiently as possible. Economist Zoë Hitzig argues that matching algorithms' emphasis on efficiently satisfying students' preferences assumes an \textit{equality of opportunity} principle of distributive justice, which may or may not align with the community or school district’s notions of justice \cite{Hitzig2019}. For example, equitable access to education and diverse classrooms have been central goals in SFUSD for over forty years \cite{sfusd2015}. However, even if the district were able to assign every student to their first choice school, schools would remain heavily racially and economically segregated, with students from low-income and historically marginalized backgrounds concentrated in under-served schools \cite{sfusd2018}.

Further, the only way to give additional advantages to a student within the constraints of the matching algorithm is by increasing their chances of receiving their first choice assignment. SFUSD does give priority in their assignment system to students who live in neighborhoods with the lowest performing schools \cite{policy5101}. While this has allowed the majority of eligible students to attend their first choice school, many of these students continue to enroll in under-served schools \cite{sfusd2018}. As we discussed in the previous section, even though all students can theoretically apply to any school, not all students can or want to attend the highest performing schools for various social and economic reasons. Therefore, equality of opportunity only exists on a surface level and is not actually achieved in practice. As a result, priority admission turns out to be a very limited avenue to promote educational equity.

Explaining the ways in which artifacts contain political properties, Langdon Winner warned of ``the all too common signs of the adaptation of human ends to technical means'' \cite{Winner1980}. Preference aggregation procedures constrain human ends to their technical means in two respects. First, participation is limited to submitting preferences in whatever form the system accepts. As we have established, certain forms of preference elicitation may be particularly subject to social biases and stereotypes or may offer unequal opportunities for families to express their needs. Second, social outcomes are limited to this procedure's range of outputs given the submitted preferences. When the preferences provided as input to such a procedure reflect social biases and inequities, we should not expect the aggregation of those preferences to promote equity. Therefore, we must consider ways of expanding participation in algorithmic decision-making.

\section{Participation beyond preferences}

The limitations of preferences that we have identified do not imply that preferences are a useless or dangerous form of participation in every context. Eliciting and aggregating stakeholder preferences can be useful for building community trust and buy-in when stakeholders have relatively equal power and access to resources \cite{webuildai}. The student assignment context highlights the limitations of preference elicitation because the socio-political context is extremely complex, power imbalances exist among participants, and the stakes for each participant are very high. 

One useful way to understand the strengths and weaknesses of preferences as a form of democratic participation is by considering the ways in which preferences offer stakeholders a form of \textit{voice}\footnote{We conceive of voice in the sense of Hirschman's theory of Exit, Loyalty and Voice \cite{Hirschman1972}. When members of an organization are dissatisfied, they can choose to express their views to drive change (Voice), passively wait for improvements (Loyalty), or leave the organization altogether (Exit). Submitting a ranked list of schools offers some degree of voice because families can tell the district which schools they prefer, although they cannot explain why. They are forced to be loyal to the existing schools and only privileged families have an exit option (e.g. private school).} in governing algorithmic systems. In student assignment, ranked preference lists are a coarse, subjective, and incomplete representation of families' values and needs. Families can express their relative preferences between schools, but this requires researching those schools and families are constrained to choosing among existing schools. In this way, preferences are especially limiting for those who have fewer resources for research and for those who gain relatively less value from the existing options. These problems call for more expressive and accessible forms of participation. For example, student assignment systems could be expanded to support:

\begin{itemize}
    \item \textbf{Alternative formats for expressing preferences:} Rather than explicitly ranking schools, families might report their weighted priorities over school factors, such as proximity to their home or language programs. It might be possible to choose school factors in a way that minimizes the extent to which priorities reflect stereotypes and biases.\\
    \item \textbf{More opportunities to express preferences:} In addition to expressing preferences over the \textit{output} of a system, stakeholders could also submit their preferences over its \textit{parameters}, as in \cite{webuildai}. For example, families could provide input on the order of priority categories, like sibling or neighborhood priority. \\
    \item \textbf{Public discourse and deliberation:} Matching algorithms are designed so that each individual can report their self-interested preferences without worrying about others'. More meaningful engagement in deliberation and discourse may allow families to more deeply understand each others' needs and address conflicts. Changes to the system's design should also be informed and shaped by community deliberation. For instance, if families were to submit weighted priorities over school factors rather than ranking schools, community deliberation would be crucial to determining which school factors are available. \\
\end{itemize}

Expanding participation in the design and governance of algorithmic policy will require developing tools and infrastructure to support informed deliberation. In SFUSD, technical decision-support tools have proven instrumental in policy-making. In 2014, Matt Kasman, then a doctoral candidate at Stanford, presented simulations to district decision-makers showing that a change to the order of the priority categories would inhibit progress towards equitable access to education \cite{sfusd2015}. These results influenced policymakers to reject a proposal to give local students higher priority than those in underserved neighborhoods. This example demonstrates the potential for technical tools to support informed policy-making, but significant further work is needed to expand these tools to enable other stakeholders, such as families, to guide analyses, interpret results and deliberate decisions. This will require making relevant information publicly accessible, developing a common language for participants to communicate about technical possibilities and constraints, and conveying how changes to the system might affect outcomes of interest. Opportunities for participation should particularly provide a low-cost platform for marginalized community members to engage. This could mean minimizing the time required to participate, and addressing participation barriers such as transportation and childcare.

Student assignment is likely to remain a contentious domain, even with successful efforts to broaden participation. Richer forms of participation should recognize and accommodate conflicts, with awareness of and sensitivity to historical patterns of power and exclusion \cite{Fraser1990, Mouffe1999}. Crawford has warned of the danger of understanding algorithmic logics as autocratic \cite{Crawford2016}. SFUSD's student assignment algorithm shows that even apparently democratic forms of participation can result in autocratic algorithmic decision-making in which it can be difficult to intervene when problems arise. Ascribing too much significance to individual preferences has created only a facade of democracy while limiting stakeholders' ability to contest algorithmic logics, challenge existing inequities, and build better futures.


\bibliography{refs}
\bibliographystyle{icml2020}

\end{document}